\documentclass[final,5p,times,twocolumn]{elsarticle}

\usepackage[utf8]{inputenc}
\usepackage{orcidlink}

\usepackage[english]{babel}
\usepackage{amsmath}
\usepackage{amsfonts}
\usepackage{amssymb}
\usepackage{graphicx}
\usepackage{xcolor}
\usepackage{hyperref}
\usepackage{epstopdf}
\usepackage{pdfpages}
\usepackage{placeins} 
\usepackage{booktabs}
\usepackage[super]{nth}
\usepackage[normalem]{ulem} 
\usepackage{xspace} 
\usepackage{caption}
\usepackage{subcaption}
\usepackage{mathtools} 
\usepackage{pdflscape}
\usepackage{multirow}
\usepackage{makecell}

\usepackage{cleveref}

\definecolor{amethyst}{rgb}{0.6, 0.4, 0.8}
\definecolor{green}{rgb}{0.55, 0.71, 0.0}
\definecolor{apricot}{rgb}{0.98, 0.81, 0.69}
\definecolor{auburn}{rgb}{0.43, 0.21,0.1}
\definecolor{babyblueeyes}{rgb}{0.63, 0.79, 0.95}
\definecolor{bittersweet}{rgb}{1.0, 0.44, 0.37}
\definecolor{blue(munsell)}{rgb}{0.0, 0.5, 0.69}
\definecolor{oceanboatblue}{rgb}{0.0, 0.47, 0.75}
\definecolor{brightmaroon}{rgb}{0.76, 0.13, 0.28}
\definecolor{deepcarminepink}{rgb}{0.94, 0.19, 0.22}

\newcommand\arcdeg{\mbox{$^\circ$}\xspace}%
\newcommand\degr{\arcdeg}%

\begin{document}

\begin{frontmatter}

\title{Direct comparison of SiPMs and PMTs in operation with a bright background and prospects of using SPADs as truly digital sensors}

\author[MPI_address]{Razmik Mirzoyan\corref{mycorrespondingauthor}\orcidlink{0000-0003-0163-7233}}
\ead{razmik@mpp.mpg.de}

\author[MPI_address]{Alexander Hahn\orcidlink{0000-0003-0827-5642}}

\author[MPI_address]{David Fink}

\author[MPI_address]{Antonios Dettlaff}

\author[MPI_address,ICRR_address]{Daniel Mazin\orcidlink{0000-0002-2010-4005}}

\author[MPI_address]{David Paneque\orcidlink{0000-0002-2830-0502}}

\author[MPI_address]{Olaf Reimann}

\author[MPI_address]{Thomas Schweizer}

\author[MPI_address]{Derek Strom\orcidlink{0000-0003-2108-3311}}

\author[MPI_address,ICRR_address]{Masahiro Teshima}

\author[MPI_address]{Yazhou Zhao}
 
\cortext[mycorrespondingauthor]{Corresponding author}

\address[MPI_address]{Max Planck Institute for Physics (Werner Heisenberg Institute), Boltzmannstr. 8, 85748 Garching, Germany}
\address[ICRR_address]{Institute for Cosmic Ray Research, The University of Tokyo, 5 Chome-1-5 Kashiwa-no-Ha, Kashiwa, Chiba, 277-8582, Japan}

\begin{abstract}
The use of silicon photomultipliers (SiPMs) alongside conventional photomultiplier tubes (PMTs) is a remarkable technological development in modern ground-based very high energy gamma-ray astronomy. SiPMs exhibit comparable or even higher photon detection efficiencies (PDEs) than PMTs. The sensitivity of a PMT matches well the spectral shape of Cherenkov radiation from extended air showers. In contrast to a PMT, the sensitivity of a SiPM is shifted toward longer wavelengths, where the intensity of light of night sky (LoNS), considered as unwanted noise, increases significantly. It is obvious that a SiPM with a higher PDE will indeed measure more Cherenkov light than a PMT, but it will also detect significantly higher LoNS noise; the question is which factor will predominate in the signal-to-noise-ratio (SNR). To compare the performance of a PMT with that of a SiPM, we built SiPM-based modules and installed these and operated in parallel in the imaging camera of the 17 m diameter MAGIC telescope. Our long-term studies show that SiPM, despite their higher PDE, can deliver only a comparable to PMT performance. As already the name SiPM suggests, we use these semiconductor sensors analogously to classical PMTs: We amplify their small signals, digitize, and calibrate the converted amplitudes. Although SiPM is essentially a digital sensor, its common-anode design does not allow one to directly profit from it. Numerous arrays of single-photon avalanche diodes (SPADs) are being developed in various laboratories worldwide. Unlike SiPM, SPAD arrays digitize the incident photons from the outset and count their number. We will dwell on the potential further developments of SPADs.
\end{abstract}


\end{frontmatter}


\section{Introduction}
Imaging Atmospheric Cherenkov Telescopes (IACTs) detect the faint, nanosecond-fast Cherenkov light flashes from extensive air showers \textbf{(EAS)} initiated by high-energy cosmic and gamma rays in the atmosphere. The performance of such telescopes depends on the photon sensors used in the focal-plane imaging camera. Traditionally, PMTs have been used due to their high gain, fast time response, low noise and low cost per photocathode area \cite{mirzoyan_technological_2022}. However, SiPMs have seen rapid technological advance, offering competitive or even better photon detection efficiency (PDE), robustness, and lower operating voltages. This study performs an \textit{in-situ comparison} of SiPM and PMT modules within the 17\,m diameter MAGIC-I imaging Atmospheric Cherenkov Telescope \textbf{(IACT)}, under real observation conditions. The original results of the comprehensive study and comparison of PMT and SiPM sensors, used over nine years in MAGIC-I, including all important sensor characteristics such as the time resolution, calibration, zenith distance dependence, etc., were published in \cite{hahn_direct_2024}. In this report, we briefly summarize the SNR achievable with both sensor types and explain the main differences when using the sensors in the presence of an intense light background, such as the LoNS. Furthermore, we briefly discuss SPAD arrays as future sensors for the construction of various imaging cameras, particularly for imaging cameras used in IACTs.

PMTs are vacuum tube devices with high gain ($4\times10^4$ -- $10^7$), single photo electron (ph.e.) response, low noise and dark counts, use high voltage and since the mid 1930's successfully used in experimental physics \cite[e.g.][]{renker_new_2009,kubetsky_multiple_1937}. The size of PMT photo cathodes can be chosen from a few to 500\,mm. SiPMs are modern solid-state detectors comprising an array of single photon avalanche diodes (SPADs) with a common anode \cite[e.g.][]{renker_advances_2009,bondarenko_limited_2000,p_buzhan_advanced_2002,buzhan_silicon_2003,dolgoshein_status_2006}. Today they can offer potentially higher PDE than PMTs, insensitivity to magnetic fields and exposure to bright light, low operational voltage and power and mechanical robustness. Largest size of a SiPM chip is limited to $\leq$6--7\,mm.

\section{Performance Metrics of an IACT}

Key performance characteristics for IACT operations are:
\begin{itemize}
\vspace{-1mm}
\linespread{0.9}\normalsize 
\setlength\itemsep{-1mm}
    \item Shape of the PDE curve of a sensor compared to the spectrum of Cherenkov light from EAS under varying observational angles.
    \item Signal-to-noise ratio (SNR), properly taking into account the unavoidable integration of LoNS noise and cross-talk \cite{dolgoshein_large_2012}.
    \item Size and reflectance of the reflector.
    \item Time response of the sensors, of the reflector and of the telescope as a whole.
    \item Dynamic range and saturation both for the light sensors and for detecting energy-dependent EAS.
    \item Threshold in number of ph.e.s and the lowest energy that can be reliably detected .
    \item Slewing time - this is important for the early follow-up of alerts of transient events in the sky.
    \item Operational robustness, also under varying LoNS and ambient light conditions.
\end{itemize}

Due to SiPM’s high sensitivity at longer wavelengths, they measure higher level of LoNS than PMTs, which degrades their SNR. This, along with the cross-talk effect are the key issues in evaluating their performance against that of PMTs.

\subsection{The telescope and camera layout}
The MAGIC telescope system comprises two $17\,\mathrm{m}$ IACTs located $85\,\mathrm{m}$ apart at the Roque de los Muchachos Observatory on the Canary Island of La Palma, at $2200\,\mathrm{m}$ a.s.l. Each telescope has an imaging camera with 1039 PMT-based channels at the focal plane of $17\,\mathrm{m}$. Every 7-channels are electro-mechanically assembled into a module of hexagonal configuration; thus, the entire camera consists of 169\,modules \cite{magic_collaboration_major_2016}. These are arranged in a circular shape on the hexagonal-shape mechanical (and temperature stabilization) support structure. This design allows for the insertion of up to six additional modules into unused vertices of the hexagon.

\subsection{SiPM Module Prototypes}
\begin{figure}
    \centering
    \includegraphics[width=1\columnwidth]{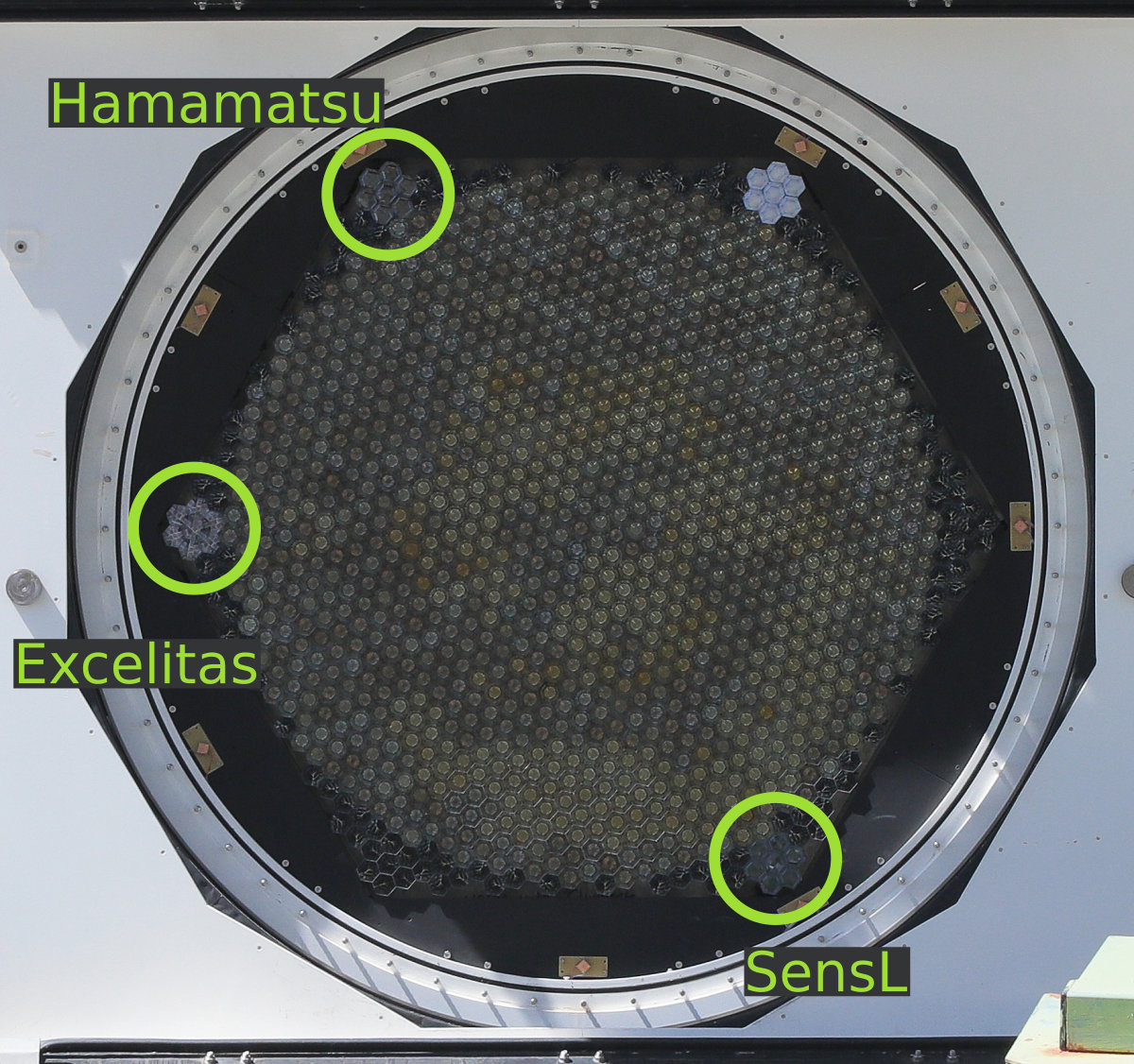}
    \caption{3 SiPM modules based on SiPMs from Hamamatsu, EXCELITAS and SensL installed in the free corners of the MAGIC-I telescope imaging camera. In the top right one can see also a 4th module, which was used for test purposes. All these modules are operated along with the PMT modules in the MAGIC camera and treated in the same way. Picture taken from \cite{hahn_direct_2024}.}
    \label{fig:MAGIC_camera}
\end{figure}

Prototype SiPM modules were designed to be optically, mechanically and electronically compatible with the existing PMT camera and modules. Each module is composed of 7 composite pixels, where each pixel consists of a closely-packed matrix of 9 (or 7, for EXCELITAS) individual SiPM chips, combined to match the equivalent sensitive area of the PMT that is coupled to the exit of used light guides. When multiple SiPM chips are connected in parallel, their terminal capacitances add up, resulting in a slow response time. This configuration leads to significant noise integrated from LoNS. Even if the pulse shape is processed and by using special filters is converted to a narrow pulse in the electronic chain, the high noise due to LoNS is already integrated. This "dilutes" small signals, leading to a high detection threshold and large fluctuations. Therefore, this solution is far from optimum and should be avoided. In order not to compromise the signal bandwidth and not to slow down the response speed of a composite, large size pixel, we have developed an active summing scheme. We used common-base transistor stages to decouple the individual capacitances and to sum up the signal from 9 (or 7) SiPM chips. Details are described in \cite{fink_sipm_2016}. The pulse shape of such a composite pixel design only marginally differs from that of a single chip \cite{hahn_direct_2024}. To compare sensors from different manufacturers and designs we built 3 types of modules, based on SiPMs from the companies Hamamatsu, SensL and EXCELITAS, see \cref{fig:MAGIC_camera}.

\section{Sensor Characteristics and Spectral Response}

\begin{figure}
    \centering
    \includegraphics[width=1\columnwidth]{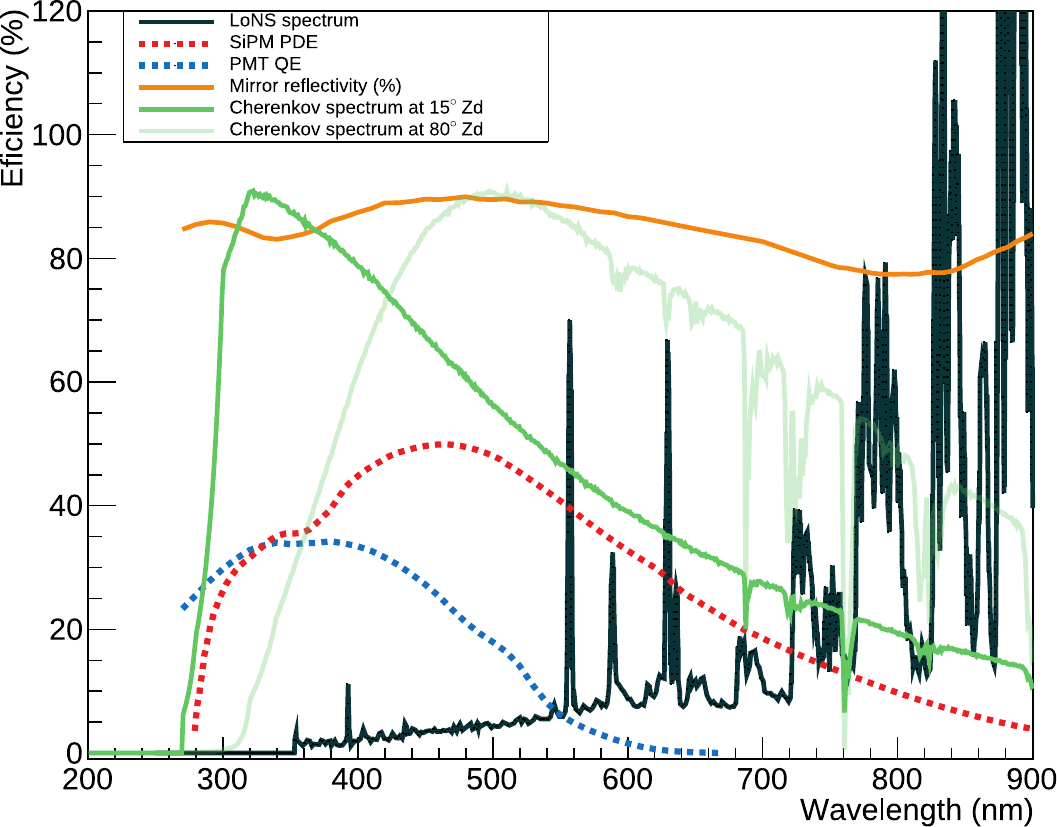}
    \caption{PDE of Hamamatsu SiPM shown in red dashed-line, QE of MAGIC PMT shown in blue, LoNS spectrum (black, in arbitrary units (A.U.)) and Cherenkov spectra of air showers observed under $15\degr$ (green, in A.U.) and $80\degr$ (light green, in A.U.) zenith distance. The mirror reflectivity of the MAGIC-1 mirrors is shown in orange. Cherenkov and LoNS spectra were scaled for better visibility. Picture taken from \cite{hahn_direct_2024}.}
    \label{fig:LoNS_response}
\end{figure}

Photon detection efficiency is a key parameter for Cherenkov light detection. PMTs used in MAGIC employ bialkali type photocathodes with peak quantum efficiency in the near-ultraviolet to blue wavelength range, closely matching the Cherenkov light emission spectrum \cite{hsu_pmt_2008,boral_tridon_magic-ii_2010,magic_collaboration_major_2016}, see \cref{fig:LoNS_response}. SiPMs, in contrast, can exhibit higher PDE at longer wavelengths (in blue to red). While the latter can increase the total number of detected photons, it also leads to enhanced integration of LoNS, which is significantly more intense at longer-wavelengths \cite[e.g.][]{benn_brightness_1998}. Given such a strong background illumination, the PDE alone is not a sufficient metric for performance comparison. The signal-to-noise ratio must be properly evaluated and of course, one shall not forget to include the negative effect of cross-talk in SiPM. Monte Carlo simulations based on CORSIKA air-shower modeling \cite{heck_corsika_1998} and adapted MAGIC detector simulations \cite{moralejo_mars_2009} were used in this study to estimate the wavelength-dependent Cherenkov and LoNS photon yields and for comparing to measured experimental results.

\section{Light of Night Sky and SiPM Intrinsic Noise}

In addition to LoNS, SiPMs exhibit intrinsic noise sources such as dark counts, optical cross-talk, and afterpulsing. While these effects are largely subdominant compared to the rate produced by LoNS under standard observing conditions, they contribute to the overall noise budget and must be properly accounted for in calibration and analysis.

\section{Signal-to-Noise Ratio and Results}

\begin{figure}
    \centering
    \includegraphics[width=1\columnwidth]{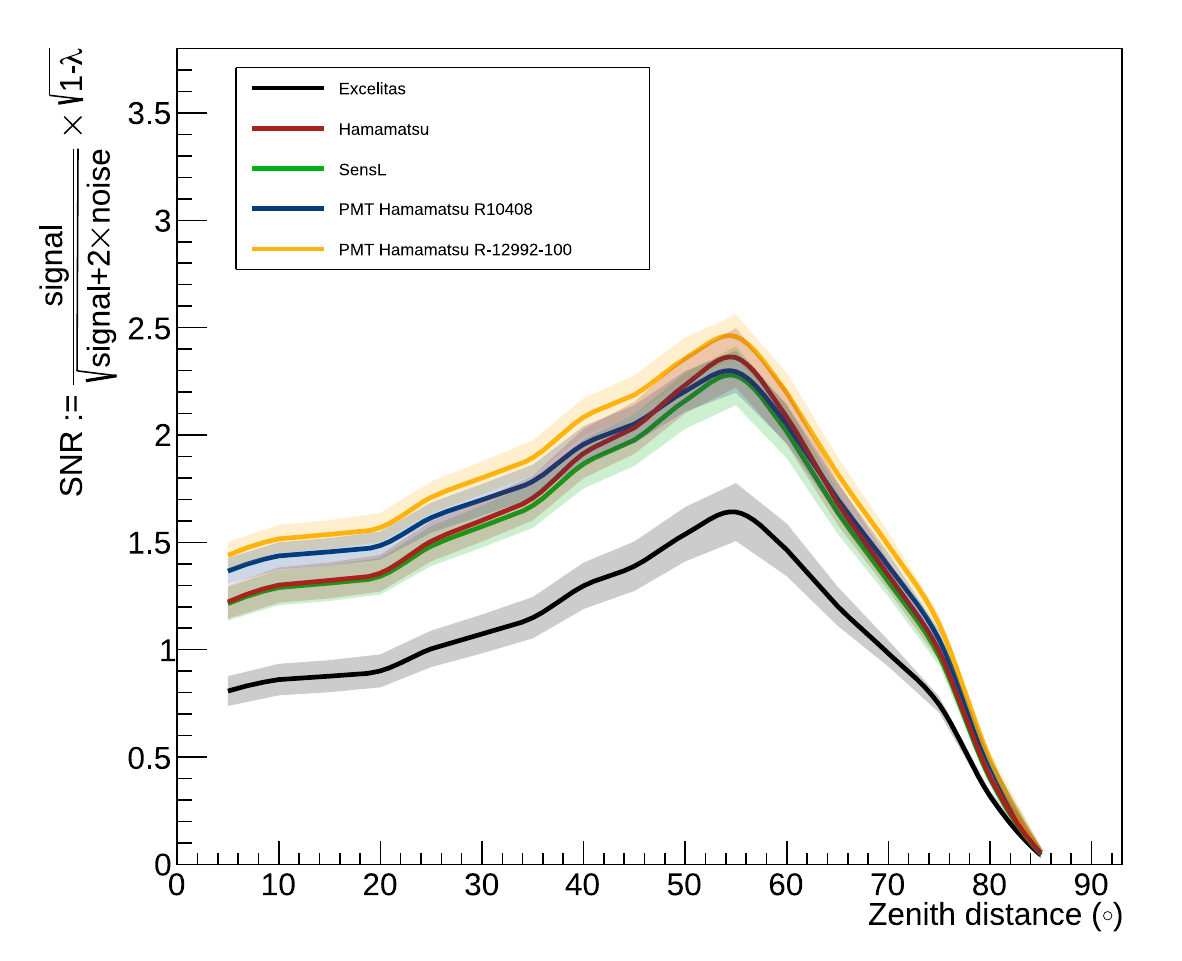}
    \caption{Dependence of the signal-to-noise ratio on the observational angle for the tested 3 SiPM modules and 2 PMT types. Except for the “old” SiPM from EXCELITAS from the year 2011, all the other sensors show comparable signal-to-noise ratios. Picture taken from \cite{hahn_direct_2024}.}
    \label{fig:SNR_vs_Zd}
\end{figure}

The signal-to-noise ratio (SNR) was evaluated by relating the expected Cherenkov signal amplitude to the standard deviation of the noise, dominated by LoNS fluctuations. The analysis accounted for the Poissonian nature of photon arrival times and the additional variance introduced by the cross-talk effect. SNR was studied as a function of zenith distance, since atmospheric absorption and shower geometry significantly influence the Cherenkov photon yield and spectral distribution of both LoNS and Cherenkov light. We found that (see \cref{fig:SNR_vs_Zd}):
\begin{itemize}
\vspace{-1mm}
\linespread{0.9}\normalsize 
\setlength\itemsep{-1mm}
    \item At low zenith distances, PMTs exhibited a slightly better SNR than SiPMs due to their superior spectral matching to the Cherenkov light emission peak and lower LoNS sensitivity.
    \item At medium and large zenith distances, the SNR achieved by SiPM modules is comparable to that of PMTs, with small differences within statistical uncertainties.
    \item Newer PMT technologies developed for next-generation Cherenkov Telescope Array (CTA) IACTs \cite{mirzoyan_evaluation_2017} generally outperform both older MAGIC PMTs and current SiPM modules, though the performance differences with SiPMs is not significant.
\end{itemize}

These results demonstrate that, under realistic observational conditions of IACTs, despite the possibly higher PDE, modern SiPMs can only provide a SNR that is comparable to PMTs.

\section{SPAD arrays as the "Non Plus Ultra" solution and advantages over the use of SiPM}

The disadvantages of using a SiPM lie in the fact that it is used as a semiconductor alternative to a PMT; that is, its small signals of a few mV for individual photoelectrons are first amplified and then digitized. Nevertheless, calibration is still required to determine the number of photoelectrons. Although SiPM is essentially a digital sensor, its common anode design does not allow one to directly profit from it. A very interesting and revolutionary alternative is offered by arrays of SPADs. Unlike SiPMs, SPAD arrays digitize incoming photons from the moment individual photons are converted into ph.e.s; subsequently, only the number of ph.e.s are counted (e.g.\ see review by \cite{cusini_historical_2022}). Due to the large amplitudes of individual ph.e. (a single ph.e. from a SPAD of size $20\,\mu \mathrm{m} \times 20\, \mu \mathrm{m}$ shows an amplitude of about $150$--$200\,\mathrm{mV}$), SPADs are insensitive to noise. With SPAD arrays that have fast counting capabilities (which is exactly what we are developing at our Max Planck Institute for Physics), the construction of an imaging camera of any size can be realized in a short time; of course, the necessary computing infrastructure with high computing power must be provided for this. We have begun developing an application-specific integrated circuit (ASIC) for reading a 4$\times$4 SPAD array. Successful demonstration will allow us to significantly increase the number of SPADs and the matrix size. We will dwell on the ultrafast response time of a typical SPAD array ($\leq 100\,\mathrm{ps}$) and our developed device in more detail in a separate report.

\section{Conclusion}

The direct comparison presented in this study shows that while PMTs show a slight advantage in SNR, also SiPMs can be used in a large IACT. Of course, scaling up to full focal plane demands detailed cost and integration studies for the very large number of needed SiPM (to replace one PMT of a large-size IACT one needs about 16 SiPMs of size $6\times6\,\mathrm{mm}^2$). Also, often researchers use special filters to reduce the SiPM sensitivity to LoNS. This necessitates careful optimization of optical filters and of trigger logic. However, this has the disadvantage that beyond the light transmission bandwidth of the filter it can reflect the SiPM's own light emission, which is generated by Geiger avalanches, back to the sensor and thus deteriorate its amplitude resolution \cite{mazzillo_electro-optical_2017}. Compared to SiPMs, SPAD arrays of high-rate capability are truly digital devices, with high noise immunity, ultra-fast timing and simplicity in design. These do not need any amplifier or amplitude to digital conversion units and calibration. SPADs offer a “Non Plus Ultra” solution for easily constructing imaging cameras of arbitrary size.

\bibliography{bibliography}

\end{document}